\documentstyle[aps,pre,preprint]{revtex}
\begin{document}
\date{\today}
%\preprint{SNUTP 97-037}
\draft
\title{Extended universality of the surface curvature \\ in equilibrium crystal
shapes}
\author{Jae Dong Noh and Doochul Kim}
\address{Department of Physics and Center for Theoretical Physics,\\
Seoul National University, Seoul 151--742, Korea}
\maketitle
\begin{abstract}
We investigate the universal property of curvatures in surface models 
which display a flat phase and a rough phase whose 
criticality is described by the Gaussian model. Earlier we derived a relation
between the Hessian of the free energy and the Gaussian coupling 
constant in the six-vertex model. Here we show its validity in a  general setting 
using renormalization group arguments. 
The general validity of the relation is confirmed numerically in the RSOS model 
by comparing the Hessian of the free energy and the Gaussian coupling constant 
in a transfer matrix finite-size-scaling study. The Hessian relation gives 
clear understanding of the universal curvature jump
at roughening transitions and facet edges and also provides an
efficient way of locating the phase boundaries. 
\end{abstract}

\pagebreak
\section{Introduction}\label{sec:1}
The theory of equilibrium crystal shapes~(ECS) is well 
established~\cite{LanL,And,Wortis,JayS}. 
Consider a macroscopic amount of solid in coexistence with and surrounded
by its own fluid phase. The shape of the solid region is obtained by 
minimizing the 
total free energy of the solid-fluid interface subject to the fixed-volume 
constraint. This leads to the Wulff construction for ECS. Especially if 
one focuses on a particular direction, say the $z$ direction, 
the crystal surface is defined by the height 
$z({\bf r})$ of the surface with respect to the position 
${\bf r}=(x,y)$ in a reference plane. 
If the surface with slope ${\bf m}$ costs a free 
energy $\sigma({\bf m})$ per unit base area, the ECS is given by~\cite{And}
\begin{equation}\label{ECS_FE}
\lambda z({\bf r}) = f(-\lambda {\bf r}) \ ,
\end{equation}
where $2\lambda$ is the pressure difference between the two
phases~\cite{LanL} and $f({\bf h})$ is the Legendre transform of 
$\sigma({\bf m})$:
\begin{equation}
f({\bf h}) = 
\min_{\{\bf m\}} \left\{\sigma({\bf m})-{\bf h}\cdot{\bf m}\right\}.
\end{equation}
Here, ${\bf h}$ is the surface-tilting field conjugate to the surface 
slope. Equation~(\ref{ECS_FE}) states that the surface free energy as a 
function of the surface-tilting field is itself the height of the surface 
from the base plane up to appropriate scaling of coordinates.

Using this connection, the thermal evolution of the equilibrium shape of a
face-centered-cubic~(fcc) crystal or a body-centered-cubic~(bcc) crystal 
has been studied through the body-centered solid-on-solid~(BCSOS) model 
which is equivalent to the six-vertex model~\cite{JayS,Bei,ShoB}. 
The surface slope and the surface-tilting field in the BCSOS model correspond 
to the polarization and the electric field in the six-vertex model,
respectively.
The six-vertex model displays several ordered phases with ferroelectric or
anti-ferroelectric order and a disordered phase~\cite{LieW,Bax}. 
The disordered phase is a critical phase. Its scaling behavior 
is described by the Gaussian model, and is parameterized by the Gaussian
coupling constant $g$ or the stiffness constant $K$~\cite{NohK}. 
At zero fields a Kosterlitz-Thouless~(KT) type roughening transition takes
place~\cite{Bax}.   Below the roughening temperature $T_R$, the system is 
ordered with zero polarization when the field is small but it becomes 
rough with non-zero polarization beyond a critical value of the 
electric field. 
This transition is in the Pokrovsky--Talapov~(PT) transition universality 
class~\cite{LieW,PokT}. The PT transition is characterized by the specific
heat exponent $\alpha=1/2$~\cite{PokT}, which implies that the free energy
scales as $f\sim |{\bf h}-{\bf h}_c|^{3/2}$. 

Equation~(\ref{ECS_FE}) enables one to identify the ordered and the disordered 
phases to facet and rounded regions in ECS, respectively. 
Below $T_R$ a facet appears surrounded by rounded vicinal surface.
The rounded region in the ECS is rough in the sense that the
height-difference correlation function behaves as
$$
\langle (z({\bf r})-z({\bf r}'))^2 \rangle \sim \frac{1}{2\pi^2 g} \ln |{\bf
r}-{\bf r}'| \ .
$$
The PT transition line corresponds to the facet edge and the crystal profile 
near the facet edge is given by $z \sim (x_\perp - x_{\perp_0})^{3/2},$  
where $x_{\perp}$ is a coordinate perpendicular to the facet edge and 
$x_{\perp_0}$ is the facet-edge position.

A measurable quantity of physical importance is the surface curvature
$\kappa\equiv \sqrt{z_{x,x}z_{y,y}-z_{x,y}{}^2}$, where the subscripts denote
partial derivatives. From Eq.~(\ref{ECS_FE}), the surface curvature is 
related to the Hessian of the free energy as
\begin{equation}\label{kappa_H}
\kappa = \lambda \sqrt{ {\rm H}\left[f({\bf h}=-\lambda{\bf r})\right]} =
\frac{\lambda}{\sqrt{{\rm H}[\sigma({\bf m})]}}\ .
\end{equation}
The Hessian of a function $F({\bf x})$ is defined by
$$
{\rm H}[F({\bf x})] = \det \left|
\begin{array}{ll} F_{x_1,x_1} & F_{x_1,x_2} \\ F_{x_2,x_1} & F_{x_2,x_2} 
\end{array} \right| \ .
$$
The second equality in Eq.~(\ref{kappa_H}) follows from the identity ${\rm
H}[f({\bf h})] {\rm H}[\sigma({\bf m})] = 1$.
It has been predicted~\cite{FisW} and measured experimentally~\cite{GalBR} 
that the curvature displays a universal jump at the roughening transition with 
discontinuity
\begin{equation}
(\Delta\kappa)_{\rm KT} = \frac{2}{\pi} \frac{\lambda d^2}{k_B T_R}\ ,
\label{jump1} 
\end{equation}
where $d$ is the distance between the crystal planes. It is also
expected~\cite{AkuAY} that, for $T<T_R$,
there is universal curvature jump at the facet edge with discontinuity
\begin{equation}
(\Delta\kappa)_{\rm PT} = \frac{1}{\pi} \frac{\lambda d^2}{k_BT} \ .\label{jump2}
\end{equation}
These universal jumps are attributed to the universal nature of the roughening
transition and the PT transition. Surface fluctuations without surface-tilting 
field are assumed to be described by the Gaussian model and the universal 
jump $(\Delta\kappa)_{\rm KT}$ is related to the jump of the stiffness constant 
at the KT transition~\cite{FisW}.  On the other hand, fluctuations of the
vicinal surface near the facet edge are described by the one-dimensional free
fermion model where the world lines of fermions are interpreted as step 
excitations in the surface~\cite{AkuAY}. 
It explains the universal jump $(\Delta \kappa)_{\rm PT}$ at the facet edge. 

In a recent paper~\cite{NohK} on the six-vertex model, an exact relation 
has been found between the Hessian of the six-vertex model free energy and the
Gaussian coupling constant $g$ in the rough phase.
The relation is given as $\bar{\rm H}[f({\bf h})]=(\frac{2}{\pi g})^2$, where
$\bar{\rm H}$ is the Hessian of the six-vertex model free energy in
units of $k_BT$ with respect to the dimensionless surface-tilting field.
When one restores the dimensions, this relation becomes
\begin{equation}\label{H_g}
{\rm H}[f({\bf h})] = \left(\frac{2d^2}{k_B T \pi g}\right)^2 .
\end{equation}
The Gaussian coupling constant determines the scaling exponents of various
correlations and controls the finite-size-scaling~(FSS) behaviors, i.e., 
a set of excitation energies $\Delta E$ of 
the transfer matrix, defined for a system with strip of width $N$, 
satisfies the FSS form
\begin{equation}
{\rm Re} (\Delta E) =
\frac{2\pi\zeta''}{N}\left(\frac{m^2}{2g}+\frac{gn^2}{2}+{\cal
N}+\bar{\cal N} \right),
\end{equation}
where $\zeta''$ is the imaginary part of the anisotropy 
factor, $({\cal N},\bar{\cal N})$ a non-negative set of integers, 
and $(m,n)$ the level index, which takes integer values~(see 
\cite{NohK,Car} for details).
If one combines Eqs.~(\ref{kappa_H}) and (\ref{H_g}), one gets
\begin{equation}\label{kappa_g6v}
\kappa = \frac{2 \lambda d^2}{\pi k_BT} \frac{1}{g}\ .
\end{equation}
It was found that $g=1$ at the KT-type roughening transition in the 
fcc $(110)$ surface and $g=2$ at the PT-type facet edges~\cite{NohK}. 
If one uses them in Eq.~(\ref{kappa_g6v}), the universal jumps in
Eqs.~(\ref{jump1}) and (\ref{jump2}) are obtained.  
Relation~(\ref{H_g}) or equivalently (\ref{kappa_g6v}) is quite general in 
the sense that it does apply to the entire rough phase as well as at the 
phase transition points of the six-vertex model.
A natural question that arises is whether such general relation is
universal, in other words, whether it holds in other model systems too. 
This paper addresses to this question, and the results are affirmative.

In Sec.~\ref{sec:2} we introduce general models for crystal surfaces
which display phase transitions between flat and rough phases, 
and present renormalization group~(RG) arguments that the Hessian of the free 
energy is the scale-invariant quantity. It yields the relation between the
Hessian of the free energy and the Gaussian coupling constant in the rough
phase. The
result is given in Eq.~(\ref{H_gg}). Combining it with the theory of the 
ECS, we obtain the relation between the surface curvature and the Gaussian 
coupling constant.
To check the general theory, we present numerical results for the 
restricted solid-on-solid~(RSOS) model in Sec.~\ref{sec:3}. 
The Gaussian coupling constants, obtained in the FSS study from
the transfer matrix spectra, is compared with the value obtained from 
the Hessian of the free energy. This confirms the validity of the 
universal relation. 
In Sec.~\ref{sec:4}, we discuss implications of the results and give 
brief summary. 

\section{Renormalization Group Theory}\label{sec:2}
Consider a solid-on-solid~(SOS) type model for a two-dimensional crystal 
surface, where the surface is defined by height $z_i$ at each site 
$i$ on a substrate of size $L_1 \times L_2$ parallel to one of
its crystal planes. The sites consist of projections of all lattice 
points on to the substrate and form a
two-dimensional lattice. Figure~1 shows three examples of such substrates for
sc (001), bcc (001), and sc (111) surfaces, respectively.
In SOS type model the height is a single-valued function; there are no
overhangs. The height at a given site can change by an integer multiple
of the lattice constant $a_3$ in the $z$ direction. 
Due to the crystal structure there may be $p$ distinct classes
 of crystal planes parallel to the substrate, with inter-plane spacing 
$d=a_3/p$. 
In that case substrate sites are separated into $p$ sublattices and 
$z_i$ takes the values $(l_i d+\mbox{integer} \times a_3)$ 
if the site $i$ belongs to the $l_i$th sublattice ($l_i=1,2,\ldots, p$).
For examples shown in Fig.~1, $p=1,~p=2$, and $p=3$ in Fig.~1 (a), (b),
and (c), respectively. Fcc (110) surfaces also have $p=2$.

At low temperatures the surface will be in a flat phase.
Steps which separate domains of flat regions are the basic excitations and
thermodynamic properties of the surface are described by a general Hamiltonian
${\cal H}$ which consists of the step-creation energy~$({\cal H}_{\rm S})$,
the interaction energy between steps~$({\cal H}_{\rm I})$, and the
surface-tilting energy~$({\cal H}_{\rm T})$ which controls the average slope
of the surface. They are given by
\begin{eqnarray}
{\cal H}_{\rm S} &=& J \sum_{\langle i,j \rangle}|z_i -z_j |^2  \label{bare_H}\\
{\cal H}_{\rm T} &=& -h_1 (\Delta z)_1 L_2 - h_2 (\Delta z)_2 L_1 \ ,
\label{bare_HT}
\end{eqnarray}
where $\langle i,j \rangle$ denotes the pair of nearest neighbor sites,
$J$ is the step energy, 
$(\Delta z)_1~((\Delta z)_2)$ is the total difference in  height across the
lattice in the $1~(2)$ direction, i.e., the difference in height between two
sites $i=(L_1,y)$ and $i=(0,y)$~($i=(x,L_2)$ and $i=(x,0)$) of the substrate,
and ${\bf h}=(h_1,h_2)$ is the surface-tilting field. 
The explicit form of ${\cal H}_{\rm I}$ is not important in the following
analysis. Since long-wavelength fluctuations are dominant in the rough
surface, we neglect height fluctuations inside the unit cell of the
two-dimensional lattice and introduce a coarse-grained height $\bar{z}_j$
which is the average height inside the unit cell containing sites $i$
around the site $j$. (The coarse-graining scheme for the sc (111) surface is
given in Ref.~\cite{NohK2}.) Then $\{\bar{z}_j\}$ takes integer multiples of
$d$.

The free energy as a function of the surface slope ${\bf m}=(m_1,m_2)$ per 
unit base area is given by
\begin{equation}\label{f1}
\sigma({\bf m}) = -\frac{k_BT}{(L_1L_2)} \ln {\cal Z}
\end{equation}
with the partition function 
${\cal Z} = 
\sum_{\{\bar{z}_i\}}' e^{-\beta ({\cal H}_{\rm S}+{\cal H}_{\rm I})}, $
where the prime denotes the sum over all surface configurations satisfying the
shifted boundary condition~(SBC) $(\Delta {\bar z})_i = m_i L_i$ $(i=1,2)$.
{\bf m} and {\bf h} are related as ${\bf h}=\nabla_{\bf m} \sigma({\bf m})$. 
Using the Poisson sum formula
$$ \sum_{n\in \delta {\bf Z}} F(n) = \sum_{n\in {\bf Z}} 
\int_{-\infty}^{\infty} \frac{d\phi}{\delta} ~ F(\phi) ~ 
e^{2\pi i n \phi/\delta}, $$
one can replace the discrete sum over $\bar{z}$ by integral over 
continuous field $\phi$ with additional harmonic terms. 
The partition function in the continuum limit is then put in the form
\begin{equation}\label{zformal}
{\cal Z} = \int_{SBC} [{\cal D}\phi] ~ 
e^{-\beta {\cal H}_{\rm eff}[\phi(\bf r)]} \ ,
\end{equation}
where the effective Hamiltonian is given by
\begin{equation}\label{H_Vil}
\beta {\cal H}_{\rm eff} = \frac{K}{2} \int d^2{\bf r}~ |\nabla
\phi|^2 + \beta {\cal H}'
\end{equation}
with
\begin{equation}
\beta {\cal H}' = -\sum_n V_n \int d^2{\bf r} \cos\left(\frac{2\pi n \phi({\bf
r})}{d}\right) + \beta {\cal H}_{\rm I}[\phi({\bf r})] \ .
\end{equation}
The functional integral in Eq.~(\ref{zformal}) is taken over the field
satisfying the SBC
\begin{equation}\label{SBC}
\phi({\bf r} + L_i {\bf e}_i) = \phi({\bf r}) + m_i L_i ,
\end{equation}
where ${\bf e}_1~({\bf e}_2)$ is the unit vector in the $1~(2)$ direction.
Here we assume for simplicity that the substrate is a square lattice whose
lattice constants in $1$ and $2$ directions are the same so that the
stiffness constant $K=2\beta J$ is a scalar. We will discuss later the more
general case where the stiffness constant is a tensor. The sine-Gordon~(SG)
terms $V_n \cos(2\pi n \phi/d)$ account for the discreteness of heights.

The effective Hamiltonian ${\cal H}_{\rm eff}$ is the starting point of our RG
arguments for the universal relation between the Hessian of the  free energy
and the Gaussian coupling constant. 
The RG theory for ${\cal H}_{\rm eff}$ with ${\bf m}=0$ and in the absence of
${\cal H}_{\rm I}$ is well established~\cite{JosKKN,NozG,IzyS}. 
At high temperatures where the surface is rough, all SG terms are
irrelevant and hence $V_n$'s renormalize to zero and $K$ renormalizes to a
fixed-point value $K^\star$. And at low temperatures the leading harmonics
becomes relevant and the surface is flat. At the roughening transition
$K^\star$ takes the universal value of $\frac{\pi}{2}$.
The RG transformation for ${\cal H}_{\rm eff}$ in the presence of ${\cal H}_{\rm
I}$ and nonzero ${\bf m}$ can be performed similarly in the following way: (i)
First introduce a field variable $\phi_0({\bf r}) \equiv \phi({\bf r})-{\bf
m}\cdot{\bf r}$, which satisfies the periodic boundary condition~(PBC) 
$\phi_0({\bf r}+L_i {\bf e}_i) = \phi_0({\bf r})$. The partition function is
then re-written as a functional integral over the field $\phi_0$ as
$$
Z = e^{-\frac{1}{2} K|{\bf m}|^2L_1 L_2} \int_{PBC} [{\cal
D}\phi_0]\  \exp\left\{ -\frac{K}{2} \int d^2 \ {\bf r} |\nabla \phi_0|^2 -
\beta {\cal H}'[\phi_0({\rm r})+{\bf m}\cdot{\bf r}] \right\} \ .
$$
(ii) The field $\phi_0$ is expended in a Fourier integral as
$$
\phi_0({\bf r}) = \int_{|{\bf p}|<\Lambda} \frac{d^2{\bf
p}}{(2\pi)^2} \ {\tilde \phi}_0({\bf p}) \ e^{i {\bf p}\cdot{\bf r}} \ , 
$$
where $\Lambda$ is the ultraviolet cutoff, and is separated into two parts
$\phi_0'({\bf r})$ and $\phi_0''({\bf r})$ such that 
$\phi_0({\bf r})=\phi_0'({\bf r})+\phi_0''({\bf r})$ and $\phi_0'~(\phi_0'')$
has only $0<|{\bf p}|<\Lambda'~(\Lambda'<|{\bf p}|<\Lambda)$ components of the
Fourier modes.
Then the partition function is decomposed as 
\begin{eqnarray*}
Z &=& e^{-\frac{1}{2}K |{\bf m}|^2 L_1 L_2} 
\int_{PBC} [{\cal D}\phi_0'] e^{-\frac{K}{2} \int d^2{\bf r} \ |\nabla
\phi_0'|^2 } \times \\
&& \int_{PBC}[{\cal D}\phi_0''] \exp \left\{ -\frac{K}{2} \int d^2{\bf r}|\nabla
\phi_0''|^2 - \beta{\cal H}'[\phi_0'({\bf r})+{\bf m}\cdot{\bf
r}+\phi_0''({\bf r})]\right\} \ .
\end{eqnarray*}
(iii) A partial integration over the fluctuations of $\phi_0''$ is performed
and the remaining field $\phi_0'$ is transformed back to $\phi'({\bf
r})\equiv\phi_0'({\bf r})+{\bf m}\cdot{\bf r}$, which corresponds to the
long-wave-length fluctuation part of $\phi({\bf r})$. 
(iv) The RG transformation is completed by rescaling the momenta or 
the coordinate and the field as
$$
{\bf p} \rightarrow b ~ {\bf p} \quad \mbox{or} \quad {\bf r} \rightarrow
{\bf r}/b \ ,
$$
\begin{equation}\label{zeta}
\phi'({\bf r}) \rightarrow \zeta \phi_{new}({\bf r}/b)
\end{equation}
with the scale factor $b=\Lambda/\Lambda'$.
The scale factor $\zeta$ for the field will be taken to be 1 to describe the
Gaussian fixed point for the rough phase. Combining Eqs.~(\ref{SBC}) and 
(\ref{zeta}), one can see that $\phi_{new}$ satisfies the SBC
\begin{equation}\label{m_tr}
\phi_{new}({\bf r}+\frac{L_i}{b}{\bf e}_i) =  \phi_{new}({\bf r}) + bm_i \left(
\frac{L_i}{b}\right),
\end{equation}
which implies that the slope is renormalized to
\begin{equation}\label{r_m}
{\bf m}' = b{\bf m} \ .
\end{equation} 

Under the RG transformation the free energy is transformed as
\begin{equation}\label{r_fe}
\sigma({\cal P},{\bf m}) = b^{-2}  \sigma({\cal P}',{\bf m}' ) +G \ ,
\end{equation}
where ${\cal P}$ denotes a set of model parameters, $G$ is the analytic
background depending on ${\cal P}$ and possibly on ${\bf m}$, and ${\cal P}'$
is the set of the renormalized model parameters. Let us focus on step (iii)
where the partial integration over $\phi_0''$ is performed, which results in
the renormalization of model parameters. However, the difference between the
${\bf m}=0$ case is that the argument of ${\cal H}'$ is replaced by $\phi_0'
\rightarrow (\phi_0' + {\bf m}\cdot{\bf r})$, which does not participate in
the integral. Therefore $K$ and the functional form of ${\cal H}'$ renormalize
in the same way as at ${\bf m}=0$, and the renormalized values of ${\cal P}'$
and $G$ are only a function of ${\cal P}$ independent of ${\bf m}$.
As a consequence, one can readily see that the Hessian of the free energy 
is the scale-invariant quantity:
\begin{equation}\label{h1_h2}
{\rm H}\left[\sigma({\bf m})\right] = {\rm H}\left[\sigma\left({\bf
m}'\right)\right].
\end{equation}

After successive applications of the RG transformation infinitely many times,
the Hamiltonian is renormalized to 
\begin{equation}\label{RGH}
\beta {\cal H}^\star = \frac{1}{2}\int d^2{\bf r} \ \ 
K^\star \left|\nabla \phi \right|^2 
\end{equation}
with the renormalized stiffness constant ${K}^\star$ provided the surface is
in the rough phase.  When the Hamiltonian is given by Eq.~(\ref{RGH}),
the slope dependent part of the free energy is easily isolated to be 
$K^\star |{\bf m}|^2/(2\beta)$ from a transformation $\phi({\bf r})
\rightarrow\phi({\bf r})- {\bf m}\cdot {\bf r}$ so that the Hessian of
$\sigma({\bf m})$ is simply given by $(K^\star/\beta)^2$.
Thus from the scale-invariant property in Eq.~(\ref{h1_h2}), the Hessian of
the original system is also given by 
\begin{equation}\label{res}
{\rm H}[\sigma({\bf m})] = \left(\frac{K^\star}{\beta}\right)^2.
\end{equation}
In general, the stiffness constant in Eq.~(\ref{H_Vil}) may be a tensor
$K_{\alpha,\beta}$ ($\alpha,\beta=1,2$). Following the same analysis, one
can easily find that ${\cal P}'$ and $G$ in Eq.~(\ref{r_fe}) do not couple to 
${\bf m}$ either and Eq.~(\ref{RGH}) is replaced by
\begin{equation}\label{H_asym}
\beta {\cal H}^\star = \frac{1}{2} \int d^2{\bf r}~ \sum_{\alpha,\beta}
K^{\star}_{\alpha,\beta} \left(\frac{\partial \phi}{\partial x_\alpha}\right)
 \left(\frac{ \partial \phi}{\partial x_\beta}\right) \ ,
\end{equation}
where $K^\star_{\alpha,\beta}$ is the fixed-point value of $K_{\alpha,\beta}$.
The Hessian of the free energy is also obtained from the scale-invariant 
property, which yields that
\begin{equation}\label{H_Kstar}
{\rm H}[\sigma({\bf m})] = \frac{\det{(K^\star_{\alpha,\beta})}}{\beta^2} \ .
\end{equation}
The stiffness constant is not a good quantity since it depends on the
scale of the field $\phi$. 
So it is convenient to use the Gaussian coupling constant $g$ which is 
defined as the coupling constant of the Gaussian model with the Hamiltonian
\begin{equation}\label{H_G}
\beta {\cal H}_G = 
\frac{g}{4\pi} \int d^2{\bf r}\ \left| \nabla \varphi \right|^2,
\end{equation}
where the periodicity of the field $\varphi$ is set to $2\pi$~\cite{FraSZ}. 
The periodicity of the field $\phi$ is $a_3$. So it is converted to $2\pi$ by 
rescaling $\varphi = 2\pi \phi/a_3$. 
After a rotation and rescale of coordinates, the Hamiltonian~(\ref{H_asym}) 
is transformed to the form of Eq.~(\ref{H_G}) with the 
Gaussian coupling constant given by
\begin{equation}\label{normalizatoin}
g = 2\pi \sqrt{\det{(K^\star_{\alpha,\beta})}} \left(\frac{a_3}{2\pi}\right)^2 .
\end{equation}
Using Eq.~(\ref{normalizatoin}) in Eq.~(\ref{H_Kstar}) 
and ${\rm H}[f({\bf h})]  =
1/{\rm H}[\sigma({\bf m})]$, one obtains that the Hessian of the free energy 
is given by the Gaussian coupling constant as
\begin{equation}\label{H_gg}
{\rm H}[f({\bf h})] = \left[\left(\frac{a_3^2}{k_BT}\right)\frac{1}{2\pi
g}\right]^2 \ .
\end{equation}
The exact result of the six-vertex model in Eq.~(\ref{H_g}) is 
recovered since $a_3=2d$ in that case.

Combining Eqs.~(\ref{kappa_H}) and (\ref{H_gg}), one finally obtains the
universal relation between the surface curvature of the ECS and the 
Gaussian coupling constant in the entire rough phase
\begin{equation} \label{kappa_g}
\kappa = \frac{2}{\pi}\frac{\lambda d^2}{k_BT}
\left(\frac{p^2}{4g}\right) ,
\end{equation}
where $a_3=pd$ is used. Equations~(\ref{H_gg}) and (\ref{kappa_g}) 
are the main results of this paper.
For fcc $(110)$ surfaces there are two equivalent crystal planes~($p=2$). 
Inserting $p=2$ in Eqs.~(\ref{H_gg}) and (\ref{kappa_g}), one reproduces the 
exact results of Eqs.~(\ref{H_g}) and (\ref{kappa_g6v}).

\section{Numerical studies of the RSOS model}\label{sec:3}
In the previous section, we presented RG arguments for the relation between
the Gaussian coupling constant and the Hessian of the free energy. It is
obtained from the observation that the Hessian of the free energy is a
scale-invariant quantity. In this section we test the 
validity of Eq.~(\ref{H_gg}) in the RSOS model on a square 
lattice~(denoted by ${\cal L}$) by comparing the Gaussian coupling constant
obtained from the FSS amplitudes of the transfer matrix spectra and the value
obtained from the Hessian of the free energy, using Eq.~(\ref{H_gg}). 

The RSOS model describes the surface of sc crystals
viewed from the $[001]$ direction.  The Hamiltonian for the RSOS model 
with the surface-tilting field ${\bf h}=(h_1,h_2)$ is given by
\begin{equation}
{\cal H}_{\rm RSOS} = K\sum_{\langle i,j\rangle} \delta\left( 
|z_i-z_j| -1 \right) - h_1\sum_i(z_{i+{\bf e}_1}-z_i) -
h_2\sum_i(z_{i+{\bf e}_2}-z_i) ,
\end{equation}
where $z_i$ is the integer-valued height variable at site $i$ in ${\cal L}$, 
$K$ is the step energy, $\langle i,j\rangle$ denotes the pair of 
nearest-neighbor sites, and ${\bf e}_1({\bf e}_2)$ is the unit vector in 
$1~(2)$ direction. The height differences between nearest-neighbor 
sites are restricted to $0$ and $\pm 1$.~(In this section, length and energy
are measured in units of lattice constant and $k_BT$, respectively. 
So all quantities are dimensionless.)
The RSOS model with ${\bf h}=0$ displays a roughening transition at $K=K_c
\sim 0.633$ and $g=1/4$ at the roughening transition~\cite{Nijs}.
There is one equivalent crystal plane parallel to the (001)
surface. This means that the RSOS model represents a $p=1$ case among the
general cases discussed in Sec.~\ref{sec:2}.

Height configurations of the RSOS model can be mapped to arrow configurations
on bonds of the dual lattice denoted by ${\cal L}_D$. 
If there is no step across a bond in ${\cal L}_D$, no arrow is assigned to
the bond. And if there is a step, an arrow is assigned in such a way that 
the height at the right-hand side of the arrow is higher than the
other side by 1.   Since there are no dislocations, the number of inward 
and outward arrows at each vertex should be equal~(the so-called ice rule). 
There are nineteen vertex configurations satisfying the ice rule. So the RSOS model
is equivalent to the 19-vertex model. The vertical~(horizontal) slope
corresponds to a net imbalance between left-right~(up-down) arrows.

A row-to-row transfer matrix ${\bf T}$ is easily constructed. 
The partition function on a lattice of size $N\times M$ can be written 
as $Z = {\rm Tr} ~{\bf T}^M$.
If one uses the PBC for the arrow variables in the $N$ direction, 
the net number $Q$ of up arrows in each row of vertical bonds is the same 
in all rows due to the ice rule. 
So the transfer matrix is separated into blocks of the form 
$$ {\bf T} = \bigoplus_Q e^{h_1 Q} ~ {\bf T}_Q ,$$ 
where ${\bf T}_Q$ operates on the $Q$th sector, defined by the set of 
arrow configurations with net number $Q$ of up arrows ($Q=-N,-N+1,\ldots,N$).
The largest eigenvalue of ${\bf T}_Q$ will be denoted by $\Lambda_Q=\exp[
-E_Q(N)]$ and $E_Q(N)$ will be called the ground-state energy in the 
sector $Q$. In the limit $N, M\rightarrow\infty$, the free energy 
$e(m_1,h_2)$ as a function of the horizontal slope $m_1=Q/N$ and the vertical 
surface-tilting field $h_2$ is given by
$$ e(m_1,h_2) = \lim_{N\rightarrow \infty} \frac{E_{m_1N}(N)}{N}\ .$$
It is related to the free energy $f({\bf h})$ 
through the Legendre transform
$$ f({\bf h}) = \min_{-1\leq m_1\leq1}\{e(m_1,h_2) - h_1m_1\}$$ 
and equilibrium values of $m_1$ and $h_1$ are related by $h_1 = \partial 
e(m_1,h_2)/\partial m_1$.

It is well known that for a rough surface with average 
horizontal slope $m_1$ the transfer matrix spectra follow the FSS form 
\begin{eqnarray}
E_{Q}(N) &=& N e(m_1,h_2) - \frac{\pi\zeta''c}{6N}, \label{eq0} \\
E_{Q\pm n}(N) &=& E_{Q} \pm n h_1(m_1,h_2) + 
\frac{2\pi\zeta''}{N}\frac{n^2g}{2}, \label{e0pm1}
\end{eqnarray}
where $Q=m_1N$, $c=1$ is the central charge for the rough phase, 
$\zeta''$ is the imaginary part of the anisotropy factor, and $g$ is the
Gaussian coupling constant~\cite{NohK}.
To obtain an estimate for $g$, one has to know the value of $\zeta''$. 
It can be obtained from Eq.~(\ref{eq0}) using the two
ground state energies $E_{Q}(N)$ and $E_{Q'}(N')$ for two values of strip
width $N$ and $N'$ chosen to satisfy the conditions $Q=m_1 N$ and $Q'=m_1 N'$.
Combining Eqs.~(\ref{eq0}) and (\ref{e0pm1}), 
one can obtain the following estimate $g_{\rm FSS}(N)$ for $g$
\begin{equation}\label{g_fss_est}
g_{\rm FSS}(N) = \frac{N(N^2-N'^2)}{12NN'} ~ 
\left(\frac{E_{Q+1}(N)+E_{Q-1}(N)-2E_{Q}(N)}{N'E_{Q}(N)-N E_{Q'}(N')}
\right).
\end{equation} 

On the other hand if the relation~(\ref{H_gg}) holds, 
it can be obtained from the relation
\begin{equation}
g_{\rm H}(N) = \frac{1}{2\pi\sqrt{{\rm H}\left[f({\bf h})\right]}}
\end{equation}
as well, 
where ${\rm H}$ is the Hessian of the free energy in dimensionless form.  
The Hessian of $f$ is directly obtained from the 
partial derivatives of $e$:
$$ {\rm H}[f({\bf h})] = -\frac{e_{h_2,h_2}}{e_{m_1 ,m_1 }}\ .$$
The partial derivatives are evaluated numerically as
\begin{eqnarray*}
\frac{\partial^2 e}{\partial m_1^2} &=& N\left(E_{Q+1}+E_{Q-1}-2E_{Q}
\right), \\
\frac{\partial^2 e}{\partial h_2^2} &=& 
\frac{\left(E_{Q}(h_2+\delta h_2)+E_{Q}(h_2-\delta h_2)-2E_{Q}(h_2) \right)}
{{N(\delta h_2)^2}} ,
\end{eqnarray*}
where we choose $\delta h_2 = 0.001$. 
This procedure gives estimates $g_{\rm H}(N)$ for $g$. 
We use the subscripts in $g$ to show how they are obtained.

Estimates for $g$ obtained in these two ways are shown in Fig.~2, where a data
point represents a pair of values $(g_{\rm H}(N),g_{\rm FSS}(N))$. Figure~2 (a) 
shows the results for $m_1 =h_2=0.0$, and $K=0.2, 0.4,$ and $0.6$. 
For $m_1 =0$, $g_{\rm FSS}(N)$ is obtained from Eq.~(\ref{g_fss_est}) by 
choosing $Q=Q'=0$ and $N'=N-1$. 
For each $K$, data shown are for the strip widths
$N= 4,5,\ldots,10$ from left to right. A similar plot is shown in Fig.~2 (b) 
for $m_1 =1/2$, $K=0.4$, and $h_2=0.1, 0.15, 0.2,0.25$, and $0.3$. For these
cases, $N'$ in Eq.~(\ref{g_fss_est}) is chosen to be $N-2$ and
the strip widths $N$ are 6,8, and 10 from left to right. In all cases, they 
converge to the same values, i.e., the data points approach the line 
$g_{\rm H}=g_{\rm FSS}$ denoted by a broken line as $N$ increases.  
The inset in Fig.~2 (a) shows the estimates 
for $m_1 = h_2=0.0$, and $K=\ln[(\sqrt{5}+1)/2]$, where 
the exact value of $g$ is known to be $1/5$ from the self-dual property 
of the RSOS model~\cite{Nijs}.
Both quantities converge excellently to the exact value. 
The fact that $g_{\rm H}$ and $g_{\rm FSS}$
converge to the same value implies that the relation~(\ref{H_gg}) holds 
in the RSOS model.  Furthermore, as can be seen 
in Fig.~2, $g_{\rm H}$ shows better
convergence than $g_{\rm FSS}$. 
In addition to the better convergence property, $g_{\rm H}$ provides a
more convenient way of estimating $g$ than the standard FSS study of the 
transfer matrix spectra. To obtain the estimate
for $\zeta''$ one should have two ground-state energies at different strip
widths with the same value of $m_1 =Q/N$. But it is difficult to find
a set of integer values of $N$ and $Q$ which gives the same value of $m_1 $.
On the other hand, to obtain $g_{\rm H}$, one needs to evaluate the largest 
eigenvalues for a single value of $N$. So the relation~(\ref{H_gg})
presents an efficient and convenient method to study the scaling behavior of 
the rough phase.

\section{Discussions and Summary}\label{sec:4}
In this paper, we introduce general models for crystal surfaces and 
derive, using the RG arguments, the relation (\ref{H_gg}) between the Gaussian
coupling constant which determines the strength of critical fluctuations of
the rough surface and the Hessian of the free energy.
Combined with the theory of the ECS, it relates the surface curvature of the
rounded region to the Gaussian coupling constant. In particular when applied 
to the phase transition point, it explains the universal curvature jump 
$(\Delta \kappa)_{\rm KT}$ at the roughening transition. The roughening
transition takes place when the SG term in Eq.~(\ref{H_Vil}) becomes 
relevant. The RG calculations for that Hamiltonian shows that 
it has a scaling dimension $x_p=p^2/(2g)$~\cite{JosKKN}, which becomes $2$
at the roughening transition, i.e., $g=g_{\rm KT}\equiv p^2/4$. So one
obtains the universal curvature jump
\begin{equation}
(\Delta \kappa)_{\rm KT} = \frac{2}{\pi}~\frac{\lambda d^2}{k_B T_R}
\frac{p^2}{4g_{\rm KT}} = \frac{2}{\pi} \frac{\lambda d^2}{k_B T_R} .
\end{equation}
As examples, the RSOS model~($p=1$), the BCSOS model~($p=2$), and the 
triangular-Ising solid-on-solid~(TISOS) model~($p=3$)~\cite{BloH} have the 
Gaussian coupling constants $g=1/4,1,\mbox{ and, } 9/4$, respectively, 
at the roughening transition points~\cite{Nijs,NohK,NohK2} 
and hence in each the universal curvature jump by the same amount.

Below the roughening transition, there appears a facet which is separated from
the rounded regions by the PT transition line. Near the PT transition
systems become extremely anisotropic~\cite{LiP} and no 
conventional RG theory has been developed for the value of $g$ at the 
transition points. Instead, the surface
near the PT transition is studied using a random walk or free-fermion
model~\cite{AkuAY}, which predicts the universal curvature jump $(\Delta
\kappa)_{\rm PT}$ in Eq.~(\ref{jump2}). 
Combining it with Eq.~(\ref{kappa_g}), one
can see that the Gaussian coupling constant should be $g_{\rm PT} = 
2 g_{\rm KT}$ at the PT transition points.  
Using that property, the PT transition point can be located accurately. 
In the RSOS model case, it is expected that $g=1/2$ at the sc (001) facet 
boundary. In Fig.~3 we present $g_{\rm FSS}(N)$ and $g_{\rm H}(N)$ for a
surface whose horizontal slope is fixed to 0~($m_1 =0$) for several values of
$h_2$ below the roughening temperature.  
The surface remains flat below a critical value of the surface-tilting field
$h_c$. Above $h_c$ the surface becomes tilted rough.
The critical value of $h_2$ can be accurately determined from the condition 
that $g=1/2$ at the transition. The insets of Figs. 3(a) and (b) show the 
estimates $h_c(N)$ for $h_c$ obtained 
from the condition $g_{\rm FSS}(N)=1/2$ and $g_{\rm H}(N)=1/2$, respectively.
Like the Gaussian coupling constant, $h_c(N)$ obtained from 
$g_{\rm H}$ have less FSS corrections than those from $g_{\rm FSS}$.
The critical value $h_c$ at $K=1.0$ is estimated as $0.30\pm 0.01$~(marked 
by arrows in the insets of Fig.~3) using polynomial fitting in
$\frac{1}{N}$, which is consistent with a value obtained from an
alternative way~\cite{alter}.

In summary, we derived the universal relation between the Hessian of the
free energy and the Gaussian coupling constant in the rough phase of
general surface model using RG arguments. 
It relates the surface curvature at the rounded region of the ECS to the 
universal quantity. Especially if it is applied to the phase transition
points, it gives a clear understanding of the universal curvature jumps. 
The validity of the relation is checked in the RSOS model numerically.
From the numerical results, it was found that the values of the Hessian have
less finite-size corrections than the scaling dimensions obtained from
the standard FSS theory. 
So, in practical points of view, this fact provides a better way 
in studying the scaling behaviors of the rough phase and the
phase transitions in crystal surfaces.

\acknowledgements
We thank M. den Nijs for critical reading of the manuscript and comments and
M.Y. Choi for discussions. This work is supported by Korea Science and 
Engineering Foundation through the
Center for Theoretical Physics, Seoul National University, by Ministry of
Education grant BSRI 96-2420 and by SNU Daewoo Research Fund.

\pagebreak
\noindent
{\bf Figure Captions}

\noindent
{\bf Figure~1} 
{Shown are the sc crystal on a substrate parallel to the (001) plane (a),
the bcc crystal on a substrate parallel to the (001) plane (b), and the sc
crystal on a substrate parallel to the (111) plane  (c). The projected
lattice points on to the substrates form 2-dimensional lattices and their
sublattice structure is shown.}

\noindent
{\bf Figure~2} 
{The Gaussian coupling constants obtained from two different methods 
at $m_1 =h_2=0$, and $K=0.2,0.4,$ and $0.6$ (a) and at $m_1 =1/2$, $K=0.4$, and
$h_2=0.1,0.15,0.2,0.25$, and $0.3$ (b) are compared. The data are
obtained from numerical diagonalizations of the transfer matrix for strip
width $N=4,5,\ldots,10$ in (a) and $N=6,8$ and $10$ in (b). 
The inset in (a) shows $g_{\rm H}$ and $g_{\rm FSS}$ at $m_1 =h_2=0$, and 
\mbox{$K=\ln[(\sqrt{5}+1)/2]$} where the exact value of
$g$ is known to be $1/5$, whose location is indicated by the arrow. They show
the converging behaviors to the exact value.
The lines are guides to eyes.}

\noindent
{\bf Figure~3}
{The Gaussian coupling constants $g_{\rm FSS}$ (a) and $g_{\rm H}$ (b) 
are shown at $m_1=0.0$ and $K=1.0>K_c$ for several values of $h_2$. 
The insets show the estimates for the critical values of $h_2$ 
which are obtained by solving $g=1/2$ numerically. The extrapolated 
values for $h_c$ are marked by arrows. The lines are guides to eyes.}


\begin{references}
\bibitem{alter}
{The facet boundary is also obtained from the first derivatives of 
the free energy, $h_{\pm} =\left. 
{\partial e}/{\partial m_1}\right|_{m_1=0^{\pm}}$, 
which give the upper and lower parts of the facet boundary as a function of
$h_2$. At the critical value of $h_2$, $h_+$ and $h_-$ merge into a single 
value, i.e., $h_+(h_c) = h_-(h_c)=0$.  
Since the RSOS model is symmetric under the change of $h_1\leftrightarrow 
h_2$, $h_c$, the critical value of $h_2$ at $h_1=0$, is equal to the 
critical value of $h_1$ at $h_2=0$. So $h_c$ can be obtained from
extrapolation of finite size data for $h_+(h_2=0)$.}
\end{references}
\end{document}